\documentclass[prd,amsmath,amssymb,aps,reprint,superscriptaddress]{revtex4-2}

\usepackage{xcolor}
\usepackage[colorlinks=true,linkcolor=blue,urlcolor=purple,citecolor=blue]{hyperref}
\usepackage{graphicx,amsmath}
\usepackage{color}
\usepackage{booktabs}
\usepackage{bm}
\usepackage{multirow}

\newcommand{\be}{\begin{equation}}
\newcommand{\ee}{\end{equation}}

\begin{document}
\title{Detectability of Nearby Binary Neutron Stars with Future sub-mHz Gravitational Wave Missions
} 
%
\author{Zhiwei Chen}\email{chenzhiwei171@mails.ucas.ac.cn}
\affiliation{National Astronomical Observatories, Chinese Academy of Sciences, 20A Datun Road, Beijing 100101, China}
\affiliation{School of Astronomy and Space Sciences, University of Chinese Academy of Sciences, 19A Yuquan Road, Beijing 100049, China}
%
\author{Youjun Lu}\email{luyj@nao.cas.cn}
\affiliation{School of Astronomy and Space Sciences, University of Chinese Academy of Sciences, 19A Yuquan Road, Beijing 100049, China}
\affiliation{National Astronomical Observatories, Chinese Academy of Sciences, 20A Datun Road, Beijing 100101, China}
\author{Yuchao Luo}
\affiliation{National Astronomical Observatories, Chinese Academy of Sciences, 20A Datun Road, Beijing 100101, China}
\affiliation{School of Astronomy and Space Sciences, University of Chinese Academy of Sciences, 19A Yuquan Road, Beijing 100049, China}
\author{Jihui Zhang}
\affiliation{National Astronomical Observatories, Chinese Academy of Sciences, 20A Datun Road, Beijing 100101, China}
\affiliation{School of Astronomy and Space Sciences, University of Chinese Academy of Sciences, 19A Yuquan Road, Beijing 100049, China}
\author{Xiao Guo}
\affiliation{Institute for Gravitational Wave Astronomy, Henan Academy of Sciences, Zhengzhou 450046, Henan, China}
\author{Jifeng Liu} 
\affiliation{National Astronomical Observatories, Chinese Academy of Sciences, 20A Datun Road, Beijing 100101, China}
\affiliation{School of Astronomy and Space Sciences, University of Chinese Academy of Sciences, 19A Yuquan Road, Beijing 100049, China}
\author{Wei-Tou Ni}
\affiliation{International Centre for Theoretical Physics Asia-Pacific, University of Chinese Academy of Sciences, Beijing 100190, China}
\affiliation{Lanzhou Center for Theoretical Physics and Key Laboratory of Theoretical Physics of Gansu Province, Lanzhou University, Lanzhou 730000, China}

\begin{abstract}
Binary neutron stars (BNSs) are one of the most important gravitational wave (GW) sources, which provide key insights to evolution of massive binary stars and nuclear physics. Beyond Laser Interferometer Space Antenna (LISA), Taiji, and Tianqin missions, proposed concepts for next generation space-based GW observatories, including LISAmax, Folkner, and eASTROD, aim to explore the sub-millihertz (mHz) to microhertz ($\mu$Hz) frequency band. Because the proposed designs substantially suppress low-frequency noise, these detectors are expected to outperform LISA, Taiji, and Tianqin in detecting eccentric Galactic BNS systems. In this paper, we estimate the detectability of nearby inspiraling BNSs using future sub-mHz GW detectors. By utilizing compact binary population synthesis simulations to generate mock BNS samples and estimate their signal-to-noise ratios (SNRs) correspondingly for each GW detector over an observation period of $5-10$\,years, we find that LISAmax may detect $\sim520–900$ Galactic BNSs, whereas Folkner and eASTROD may detect $\sim780-1370$ Galactic BNSs. Notably, LISAmax excels in detecting highly eccentric systems $(e>0.90)$ owing to its higher sensitivity at relatively higher sub-mHz frequencies. We further identify seven observed radio BNSs as viable candidates for validation, in particular J0737-3039, which reaches an SNR of $\sim 100$. The expected detection number of LMC inspiraling BNSs is about $\sim4-18$ for these sub-mHz detectors over an observation period of $5-10$\,years, while detecting inspiraling BNSs in SMC is challenging. This study highlights the significant potential of future sub-mHz GW missions in unraveling BNS formation and evolution physics.
\end{abstract}
%

\maketitle

\section{Introduction}

The first direct detection of gravitational wave (GW) emitted by stellar binary black hole mergers GW150914 by the LIGO-Virgo-KAGRA (LVK) Collaboration in 2015 marked the dawn of gravitational wave astronomy \cite{2016PhRvL.116m1102A}, opening up a new window for exploring the universe. Since then, more than 200 compact binary coalescence (CBC) GW events at high-frequency band ($10–1000$\,Hz) have been detected in the O1-O4 run ~\cite{2019PhRvX...9c1040A, 2020arXiv201014527A, 2021arXiv211103606T,2025arXiv250818080T,2025arXiv250818082T}.  Future third-generation ground-based detectors such as the Einstein Telescope \cite{Hild_2011} and Cosmic Explorer \cite{2019BAAS...51g..35R} will further enhance the detection capability in this frequency range.

In contrast, space-based gravitational wave missions such as Laser Interferometer Space Antenna (LISA) \cite{2017arXiv170200786A}, Taiji \cite{2020ResPh..1602918L}, and Tianqin \cite{2021PTEP.2021eA107M} target the millihertz (mHz) band, enabling the detection of astrophysical sources inaccessible to ground-based instruments, including massive binary black holes (MBBHs) \cite{2021hgwa.bookE..18B,2020ApJ...897...86C}, extreme mass-ratio inspirals (EMRIs) \cite{2017PhRvD..95j3012B}, and early inspiral stage of stellar-mass compact binaries \cite{2023MNRAS.522.2951Z,2025ApJ...978...61Z}. At even lower frequencies, the nanohertz (nHz) band, pulsar timing arrays (PTAs) such as NANOGrav, EPTA, PPTA, and CPTA have provided compelling evidence for the stochastic gravitational wave background generated by the inspiral of supermassive black hole binaries \cite{2023ApJ...951L...8A,2023A&A...678A..50E,2023ApJ...951L...6R, 2023RAA....23g5024X, 2025MNRAS.536.1467M}, marking the beginning of muliti-band gravitational wave astronomy.

There is a frequency gap between the mHz band of missions like LISA and the nHz band of PTAs. To bridge this gap, a series of future sub-mHz to microhertz ($\mu$Hz) gravitational wave mission concepts have been proposed \cite{2016IJMPD..2530001N}, including LISAmax \cite{2023CQGra..40s5022M}, eASTROD \cite{2013IJMPD..2241004N}, and the Folkner mission \cite{Mueller+2019BAAS}. These missions adopt longer interferometric arm designs, such as heliocentric orbits with baselines of $\sqrt{3}$\,AU, to achieve higher sensitivity in the sub-mHz frequency range. Compared with LISA, sub-mHz detectors not only partially overlap in sensitivity but also extend significantly to lower frequencies, which is expected to open up new prospects for the detection of various gravitational wave sources \cite{2016IJMPD..2530001N,2025PhRvD.112h4045W}.

Among potential gravitational wave sources, binary neutron stars (BNSs) are of particular importance. The subsequent observations of the binary neutron star merger event, GW170817, including its multiwavelength electromagnetic (EM) counterparts not only confirmed the predictions about short $\gamma$-ray burst (sGRB) with afterglow signals and kilonova produced by BNS mergers, but also marked the opening of the new era of multi-messenger astronomy \cite{2017ApJ...848L..12A, 2017ApJ...848L..13A, 2017Sci...358.1556C, 2019MNRAS.489L..91C, 2019PhRvX...9a1001A}. Notably, BNSs formed from isolated binaries may have significant eccentricities at birth due to supernova natal kicks and mass loss \cite{2025MNRAS.544L..89C,2025ApJS..280...43T}. However, when evolving to the high-frequency window ($10–1000$\,Hz) of ground-based detectors, gravitational radiation reaction will circularize their orbits to eccentricities $e\lesssim 10^{-5}$, making it difficult to study the initial eccentricity and related binary evolution processes. Although LISA, which targets the mHz band, may detect residual eccentricities of inspiraling BNSs, its sensitivity in the sub-mHz range is largely limited \cite{2020MNRAS.492.3061L,2020ApJ...892L...9A,2020ApJ...898...71B,2021MNRAS.502.5576K,2022ApJ...937..118W}. Future sub-mHz detectors, with their superior sensitivity in lower frequency bands, are expected to detect more BNS systems with significant eccentricities, providing key insights into BNS formation channels, supernova natal kick mechanisms, and binary evolution physics.

In this work, we focus on assessing the detection capability of future sub-mHz space-borne gravitational wave detectors (LISAmax, eASTROD, and Folkner) for BNS inspiral systems in the Milky Way (MW), Large Magellanic Cloud (LMC), and Small Magellanic Cloud (SMC). We first generate mock BNS inspiral samples through compact binary population synthesis (CBPS) and galactic dynamics calculations and then calculate the signal-to-noise ratio (SNR) of each mock sample for different detectors.  This paper is organized as follows: Section~\ref{sec:method} briefly introduces the methodology for generating mock BNS inspiral systems and calculating their SNR with future sub-mHz GW detectors. Section~\ref{sec:results} presents the main results of the detection capability assessment. Conclusions and discussions are given in Section~\ref{sec:con}.

\section{Methodology}
\label{sec:method}

In this section, we briefly introduce the methodology on generating the mock BNS inspiral systems (subsection~\ref{pop}) and calculating their SNR with future sub-mHz GW detector (subsection~\ref{gw}). 

\subsection{BNS Inspiral Population}
\label{pop}

 We utilize an updated version of CBPS code \texttt{BSE} to simulate the formation of mock BNS insprial systems, which is originally developed by \cite{2000MNRAS.315..543H,2002MNRAS.329..897H}, and later modified and extended by \cite{2010A&A...521A..85Y}, \cite{2022MNRAS.509.1557C,2025ApJ...980..181C}, and \cite{2025MNRAS.544L..89C}.

The initial condition of the simulation are set as follows.  The initial mass spectrum for main-sequence (MS) binary stars are generated by assuming that, the primary mass $M_1$ follows the initial mass function given by \cite{2001MNRAS.322..231K}, i.e., $p(M_1)\propto M_1^{-2.3}$ for $M_1>1M_{\odot}$, and the mass ratio $q=M_2/M_1$ follows a uniform distribution within $(0.01,1)$ \cite{1989ApJ...347..998E}. The initial separation $a$ between binaries is assumed to follow the distribution given by \cite{1998MNRAS.296.1019H}, 
\begin{equation}
    an(a)=
    \begin{cases}
    \alpha_{\rm sep}(\frac{a}{a_0})^{1.2} & a\leq a_0 \\
    \alpha_{\rm sep} & a_0<a<a_1,
    \end{cases}
\end{equation}
where $n(a)$ denotes for the probability of binaries with semi-major axis in the range from $a$ to $a+da$, $\alpha_{\rm sep}\sim 0.070$, $a_0=10R_{\odot}$, $a_1=5.75\times10^6 R_{\odot}$. Notably, we further assume all MS binaries are initially on circular orbits and therefore possessing initial eccentricities $e_i=0$. As for the newborn NS, we assume the equation of state (EOS) to be SLy with maximum mass for non-rotating neutron stars of $M_{\rm TOV}\sim 2.06M_{\odot}$ \cite{2001A&A...380..151D} {for it is compatible with the current constraint from both GW multi-messenger signals of GW170817 and heavy pulsars \cite{2017ApJ...850L..19M,2025arXiv251022290W}. We also test that adopting a rather stiffer EOS such as DD2 \cite{2010NuPhA.837..210H} would not alter our final estimation significantly. This is because the EOS only alters the radii and the high-mass end truncation of the NS components in CBPS, while the overall NS mass distribution is determined directly by the supernova remnant mass function.
} 
In addition, we set the initial spin period $P_0=0.01$\,s. The initial magnetic field follows the log-normal distribution with the mean of $13$ and the scatter of $0.55$ for NS with progenitor MS mass $\lesssim 20 M_{\odot}$ \cite{2006ApJ...643..332F}, and uniform distribution within the range of $[10^{14},10^{15}] $\,G for NS with progenitor MS mass $\gtrsim 20 M_{\odot}$ \cite{2014ApJS..212....6O}.

With the initial condition discussed above, we simulate several important astrophysical processes in binary evolution, including the most uncertain common envelope (CE) ejection \cite{2018MNRAS.481.4009V,2018MNRAS.480.2011G,2019MNRAS.486.3213A} and  supernova explosion \cite{2005MNRAS.360..974H,2012ApJ...749...91F,2018MNRAS.481.4009V}. For brevity, we only list several key parameter settings in our CBPS simulation in Table~\ref{tab:t1} and refer the readers to \cite{2022MNRAS.509.1557C,2025ApJ...980..181C}, and \cite{2025MNRAS.544L..89C} for more detailed introduction. Notably, these chosen settings are found to be most compatible with both latest constraint on the merger rate density of BNS mergers from the preliminary catalog of full O4 LVK observation \cite{2025arXiv250708778A} and eccentricity-orbital period distribution of Galactic BNS systems observed by radio telescopes \cite{2025MNRAS.544L..89C}. After the formation (denoted as $t_{\rm f}$), the orbit of BNS will decay and circularize due to the GW radiation. The evolution of semi-major axes $(a)$ and eccentricities $(e)$ can be estimated by \cite{1964PhRv..136.1224P}:
\begin{equation}
\bigg\langle\frac{da}{dt}\bigg\rangle=-\frac{64}{5}\frac{G^3m_1m_2(m_1+m_2)}{c^5a^3(1-e^2)^{\frac{7}{2}}}\bigg(1+\frac{73}{24}e^2+\frac{37}{96}e^4\bigg)
\label{eq:dadt}
\end{equation}
and 
\begin{equation}
\bigg\langle\frac{de}{dt}\bigg\rangle=-\frac{304}{15}e\frac{G^3m_1m_2(m_1+m_2)}{c^5a^4(1-e^2)^{\frac{5}{2}}}\bigg(1+\frac{121}{304}e^2\bigg),
\label{eq:dedt}
\end{equation}
where $m_1$ and $m_2$ denotes the primary and secondary component masses of a BNS. 

As for Milky Way (MW), we assume a constant metallicity for the MW of $Z = Z_\odot$ and a uniform star formation history (SFH) similar with the treatment in \cite{2018MNRAS.481.4009V}  to generate mock BNS inspiral systems across cosmic time and exclude those have merged within their lifetime. In our simulation, we assume in total there are $\sim 6.1\times 10^{10} M_\odot$ stellar mass in the MW, and predict that there will be $\sim 5.7\times 10^{5}$ BNS inspiral systems formed. To calculate their GW SNR, we further simulate their distance $d_{\rm L}$ to the GW detector by the following procedures \cite{2020MNRAS.494.1587C}.  Firstly, we randomly distribute them throughout the MW's disk according to the density profile \cite{2013A&A...549A.137I} and assign each mock BNS a velocity $\vec{v_{i}}$ by the local circular rotation curve with a random $10$\,km/s dispersion. Notably, during the second-born supernova explosion, the natal kick will modify the initial velocity by a vector superposition $\vec{v}=\vec{v_{i}}+\vec{v}_{\rm k,2}$, where $\vec{v}_{\rm k,2}$ is directly drawn from the BSE simulation. Then, we simulate the dynamical evolution of BNS systems in the Galactic potential employing the \texttt{galpy} package \cite{2015ApJS..216...29B}, by modeling the potential of MW by three individual components, including a Plummer sphere bulge, an exponential disc and a NFW dark matter halo, of which the fitted parameters for each component are drawn from Model III of   \cite{2013A&A...549A.137I} (see Table 3 therein for details). This allows us to track each system's trajectory from its formation epoch $t_{\rm f}$ to the present observation time, providing their final distance to the GW detector $d_{\rm L}$. 

\begin{figure}
\centering
\includegraphics[width=1.00\columnwidth]{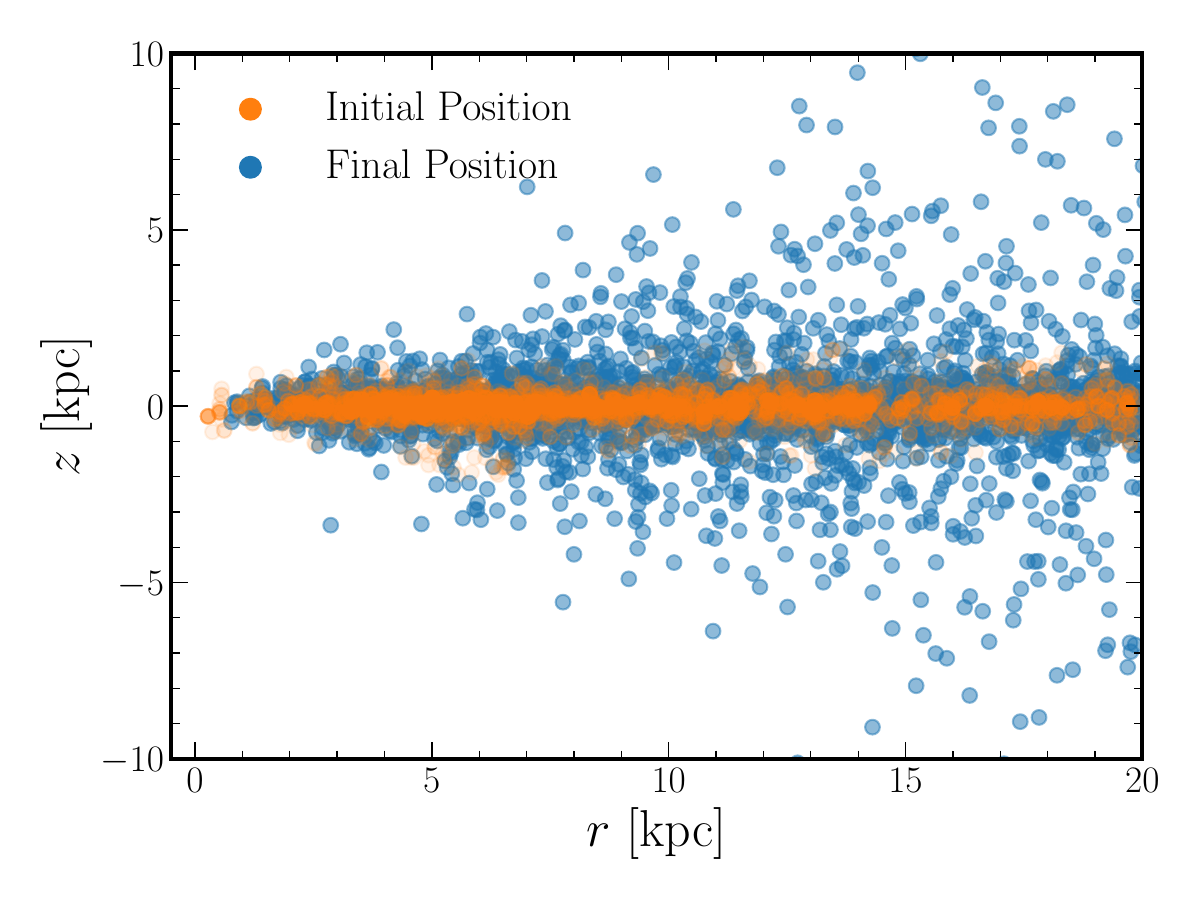}
\caption{The position $(z,r)$ of mock inspiralling BNS in MW cylindrical coordinates at the formation epoch time $t_{\rm f}$ (orange) and present observation time (blue) respectively, considering the orbital adjustment due to the second supernova explosion natal kick $\vec{v}_{\rm k,2}$. 
}
%
\label{fig:f1}
\end{figure}

\begin{figure}
\centering
\includegraphics[width=1.0\columnwidth]{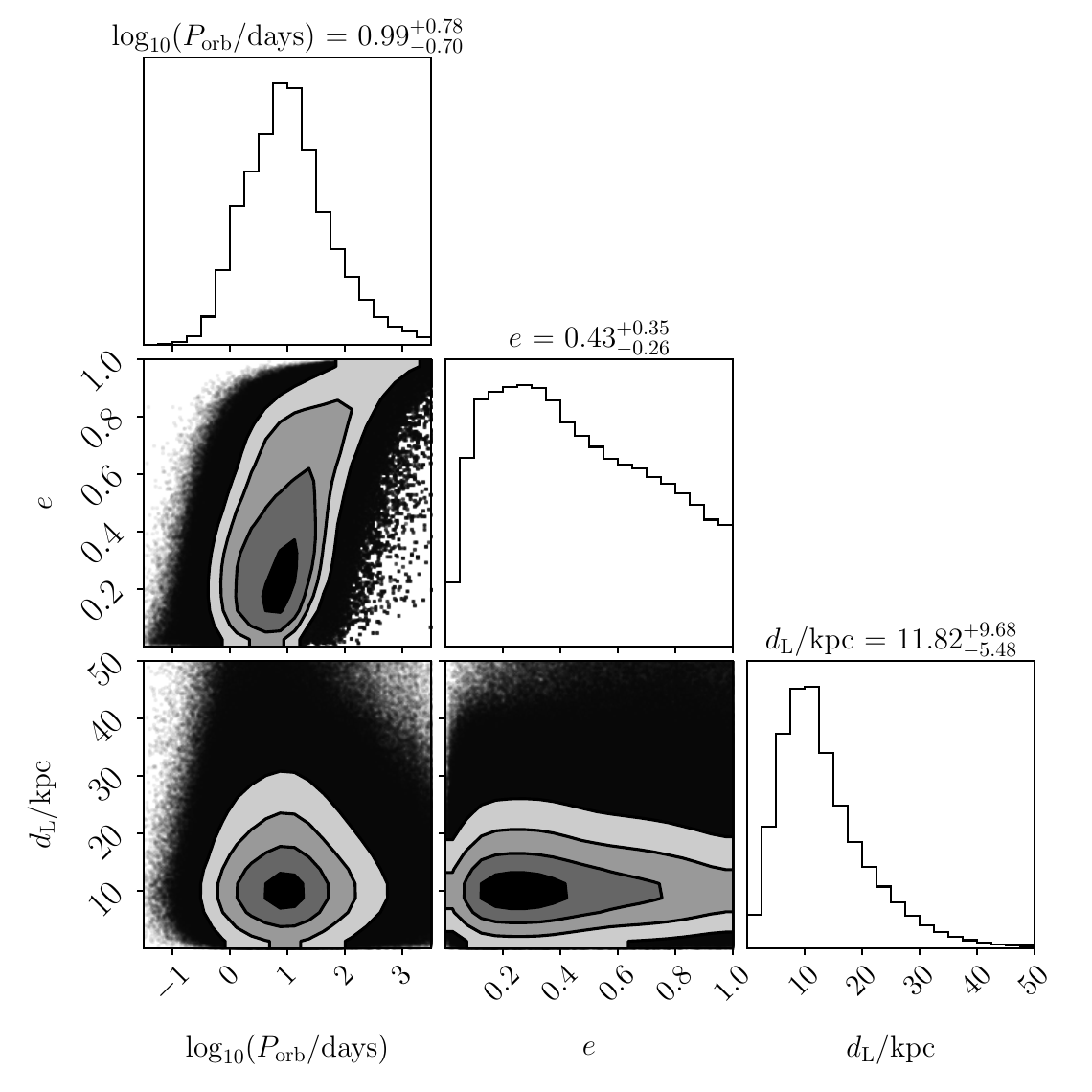}
\caption{ The joint distribution of orbital periods $P_{\rm orb}$, eccentricity $e$ and distance $d_{\rm L}$ of mock MW inspiralling BNS simulated by CBPS method.
}
%
\label{fig:f2}
\end{figure}

Figure~\ref{fig:f1} shows the position $(z,r)$ of mock inspiralling BNS in MW cylindrical coordinates at the BNS formation time $t_{\rm f}$ (orange) and present observation time (blue), respectively. The initial vertical $z$ positions of mock BNSs are limited within very small ranges $\sim \pm 1$\,kpc, i.e., located in the MW thin disk. However, their current positions are comparably more extended, with many significantly away from the MW thin disk, due to the orbital adjustment induced by random orientated supernova natal kicks. Figure~\ref{fig:f2} shows the joint distribution of orbital period $P_{\rm orb}$, eccentricity $e$ and luminosity distance $d_{\rm L}$ of mock inspiralling BNS simulated by CBPS method. The median value of $P_{\rm orb}$ is about $10$\,days and the median value of $e$ is about $0.43$. The median value of $d_{\rm L}$ is about $11.8$\,kpc (the distance from the Galactic center to the GW detector is about $8.12$\,kpc). 

Similar with MW, we assume the metallicity to be $Z=0.4Z_\odot$ and adopt a constant uniform SFH with total stellar mass $\sim 2.7\times 10^9 M_\odot$ and $\sim 3.1 \times 10^{8} M_\odot$ \cite{2015arXiv151103346B} to generate mock BNS inspiral systems in LMC and SMC respectively. Notably, the distance between LMC and SMC are significantly larger than the size of themselves. Hence, we do not simulate the dynamics of the mock samples, but rather simply assume the value of $d_{\rm L}$ of all the mock BNS inspiral systems located in LMC and SMC to be $\sim 49.7$\,kpc \cite{Walker_2011} and $\sim 62$\,kpc \cite{osti_22862851}  respectively. In our simulation, we predict that in total there will be $\sim 3.8\times 10^{4}$ and $\sim 3.8\times 10^3$  BNS inspiral systems in the LMC and SMC respectively. 

In summary, we directly extract the information on the masses $m_1$ and $m_2$, distance $d_{\rm L}$, orbital period $P_{\rm orb}$ (equivalently semi-major axis $a$) and eccentricity $e$ of mock inspiralling BNS in MW, LMC and SMC from the CBPS simulation, which is crucial for the calculation of SNR. Notably, we simulate 10 random realizations to avoid Poisson counting noise for each case. 

\begin{table}
\caption{Parameter settings for generating mock BNS inspiral systems with updated CBPS code \texttt{BSE}.  }
%
\label{tab:t1} 
\centering
\begin{tabular}{lc} 
\hline
Parameter &  Settings \\
\hline
Remnant mass function & Rapid \\
CE Structured parameter & \citet{2014AA...563A..83C}  \\
CE ejection efficiency & $\alpha_{\rm CE}=5$  \\
HG donor merger fraction & $f_{\rm HG}=0.8$ \\
CCSNe kick dispersion & $\sigma_1=265$\,km/s\\
EC/USSNe kick dispersion & $\sigma_2=30$\,km/s\\ \hline
\end{tabular}
\end{table}

\subsection{GW detection}
\label{gw}

To investigate the detectability of BNS inspiral systems with future sub-mHz GW detectors, such as  LISAmax \cite{2023CQGra..40s5022M}, eASTROD \cite{2013IJMPD..2241004N} and Folkner \cite{Mueller+2019BAAS}, we estimate the SNR of each mock sample. In our sample, most of sources will possess substantial eccentricities and thus the GW emission frequency $f_{\rm GW}$ is not simply twice the orbit frequency $f_{\rm orb}$ but a superposition of all harmonic mutiples of the orbital frequencies, i.e., $f_{\rm GW}=nf_{\rm orb}$, where $n$ is an integer larger than 0. The characteristic strain of the $n$-th harmonic for an eccentric BNS can be estimated by \cite{PhysRev.131.435},
\begin{equation}
h_{c,n}(f_{\rm orb})=\frac{1}{\pi d_{\rm L}}\sqrt{\frac{2G}{c^3}\frac{dE_{n}}{d(nf_{\rm orb})}},
\label{eq:hc}
\end{equation}
where $c$ is the light-speed, $G$ is the gravitational constant and  $f_{\rm orb}$ is the orbital frequency. The term $dE_{n}/{d(nf_{\rm orb})}$ is the GW energy emitted at the $n$-th harmonic,
\begin{equation}
\frac{dE_{n}}{d(nf_{\rm orb})}=\frac{(2\pi G)^{2/3}M_{\rm c}^{5/3}}{3f_{\rm orb}^{1/3}}\frac{g(n,e)}{nF(e)},
\end{equation}
where $M_{\rm c}$ is the chirp mass of the BNS, which is directly generated via the CBPS simulation. Here the two functions $g(n,e)$ and $F(e)$ are given by \cite{PhysRev.131.435},
\begin{eqnarray}
g(n,e)& = & \frac{n^4}{32} \left\{[J_{n-2}(ne)-2eJ_{n-1}(ne)+ \frac{2}{n}J_n(ne)+\right. \nonumber \\
& &  2eJ_{n+1}(ne)-J_{n+2}(ne)]^2+(1-e^2) [J_{n-2}(ne)  \nonumber \\
& & \left. -2J_n(ne)+J_{n+2}(ne) ]^2+\frac{4}{3n^2}[J_n(ne)]^2 \right\},
\label{eq:gne}
\end{eqnarray}
and
\begin{equation}
F(e) = \frac{1+(73/24)e^2+(37/96)e^4}{(1-e^2)^{7/2}},
\label{eq:fe}
\end{equation}
where $J_n$ is the Bessel function of the first kind.

Then the SNR of an eccentric BNS inspiral can be estimated by,
\begin{equation}
\rho^2=\sum_{n=1}^{N}\rho^2_{n}=\sum_{n=1}^{N}\int_{f_{\rm i}}^{f_{\rm f}}\frac{ h^2_{c,n}(f_{\rm orb})}{nf_{\rm orb} S_{\rm n}(nf_{\rm orb})}\frac{df_{\rm orb}}{f_{\rm orb}},
\label{eq:snr}
\end{equation}
where $S_{\rm n}(nf)$ is the two-channel sky/polarization averaged power spectral density of a GW detector. The frequency $f_{\rm i}$ and $f_{\rm f}$ are the initial and final orbital frequencies of BNS after the mission time $T_{\rm obs}$ of GW detector, which can be calculated by solving Equations~\eqref{eq:dadt} and \eqref{eq:dedt} through Runge-Kutta (RK45) method. {Note here that these two equations only describe the orbital evolution at the 2.5 Post-Newtonian (PN) order. For the long observation periods ($5-10$ years) considered here, higher-order PN corrections (e.g., 3.5PN) could introduce phase drifts that are critical for coherent matched-filtering searches. Nevertheless, since the 3.5PN term's relative contribution to the energy flux is of the order $(v/c)^2 \sim 10^{-6}$ for our typical samples, its impact on the total SNR and the total detectable number may be neglected \cite{2002LRR.....5....3B}. Therefore, we simply use these 2.5PN evolution as a demonstration for assessing the detectability of inspiralling BNS with sub-mHz detectors.} Then, we may further define the effective GW strain $h_{{\rm eff},n}$ at the $n$-th harmonic,
\begin{equation}
h_{{\rm eff},n}(f_{\rm orb})=\sqrt{\frac{(f_{\rm f}-f_{\rm i})h^2_{{\rm c},n}(f_{\rm orb})}{f_{\rm orb}}}. 
\end{equation}
This definition allows $h^2_{{\rm eff},n}(f_{\rm orb})$ to restore the average GW power emitted on each harmonic frequency interval $[nf_{\rm orb},(n+1)f_{\rm orb}]$ and a clear illustration of GW source on the characteristic strain versus frequency plot. Then if the GW frequency does not evolve much, Equation~\eqref{eq:snr} can be reduced to,
\begin{equation}
\rho^2=\sum_{n=1}^{N} \frac{h^2_{{\rm eff},n}(f_{\rm orb})}{h^2_{\rm noise}(nf_{\rm orb})},
\label{eq:rhon}
\end{equation}
where the characteristic noise strain for a GW detector is defined as $h_{\rm noise}(nf_{\rm orb})=\sqrt{nf_{\rm orb} S_{\rm n}(nf_{\rm orb})}=\sqrt{f_{\rm GW} S_{\rm n}(f_{\rm GW})}$. 
By this definition, the relative height of $h_{{\rm eff},n}$ and $h_{\rm noise}(f)$ represent the SNR of each harmonics. 

\begin{figure}
\centering
\includegraphics[width=1.0\columnwidth]{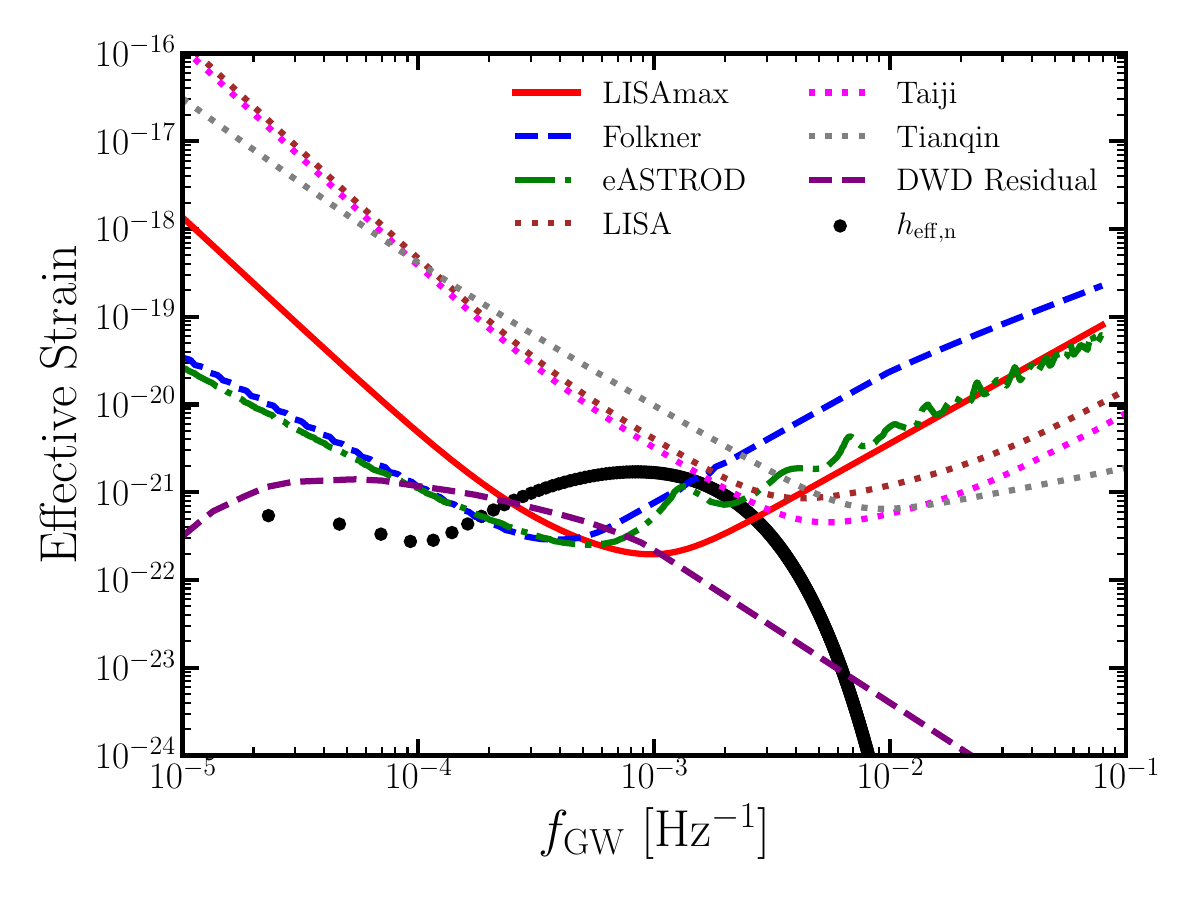}
\caption{The characteristic noise strain $h_{\rm noise}$ of future sub-mHz GW detectors with respect to GW frequency $f_{\rm GW}$, i.e., LISAmax (red solid line), Folkner (blue dashed line) and eASTROD (green dashed dotted line). The purple dashed line shows the DWD residual foreground estimated. For comparison, we also plot the $h_{\rm noise}$  of mHz detectors LISA (brown dotted), Taiji (magenta dotted) and Tianqin (grey dotted). The black dots present the effective strain $h_{\rm eff}$ at each harmonics of a BNS inspiral with $e\sim 0.90$ and $P_{\rm orb}\sim 0.5$\,days placing at a (typical) distance of $d_{\rm L}\sim 11.8$\,kpc.  
}
%
\label{fig:f3}
\end{figure}

In this work, we focus on the detection of inspiralling BNS with sub-mHz GW detectors, including LISAmax \cite{2023CQGra..40s5022M}, eASTROD \cite{2013IJMPD..2241004N}, and Folkner \cite{Mueller+2019BAAS}, and consider two different mission time for these detectors, i.e., $T_{\rm obs}=5$\,yr and $10$\,yr, respectively. Figure~\ref{fig:f3} shows the characteristic noise strain $h_{\rm noise}(f)$ of sub-mHz GW detectors and mHz detectors, such as LISA \cite{2017arXiv170200786A}, Taiji \cite{2020ResPh..1602918L} and Tianqin \cite{2021PTEP.2021eA107M} for comparison. As seen from this figure, the most sensitive frequencies of LISAmax, Folkner, and eASTROD are around $10^{-3}~\rm Hz$, which is about $\sim 10$ times better than that of mHz detectors. Moreover, the noise of sub-mHz GW detectors is significantly smaller than that of mHz detectors at frequencies inspiraling BNSs populates, indicating their substantially better capability on detection of inspiralling BNS. For illustration, we also plot the effective strain $h_{{\rm eff},n}$ of a mock BNS inspiral system with $e\sim 0.90$ and $P_{\rm orb}\sim 0.5$\,days, placing at a typical distance of $d_{\rm L}\sim 11.8$\,kpc. As seen from the figure, almost all $h_{{\rm eff},n}$ are substantially below the $h_{\rm noise}(f_{\rm GW})$ for the mHz detectors, but quite a number of $h_{\rm noise}(f_{\rm GW})$ around the peak $n_{\rm p}$ are well above those for the sub-mHz detectors, indicating it is detectable by sub-mHz detectors but not for mHz detectors. Notably, a large population of Galactic binaries, particularly double white dwarfs (DWDs), are expected to be in the early inspiral stage, continuously emitting GWs at frequencies below $\sim 10$\,mHz \cite{2017JPhCS.840a2024C,2023PhRvD.107l4022W}. The unresolved majority form a foreground or confuse noise, consequently lowering the SNR. Nevertheless, as the mHz mission progresses, the confusion noise reduces since resolved binaries may be accurately modeled \cite{2019CQGra..36j5011R,2023ApJ...945..162T}. {Thus, we follow \cite{2025PhRvD.112h4045W} and \cite{2023PhRvD.107l4022W} to optimistically add a residual noise defined by the difference between the median and $1\sigma$ uncertainty bounds of the fitting formula of the DWD foreground (see purple dashed line in Fig.~\ref{fig:f3} and Section B in \cite{2023PhRvD.107l4022W} for detailed derivation) rather than the full Galactic foreground.}

We impose the typical threshold $\rho_{\rm th}=8$ for GW detection for demonstration and estimate the number of detectable BNSs. We find that the expected detection number of the MW inspiraling BNSs with mHz detectors are about $N_{\rm MW}\sim 19_{-5}^{+7}$, $25_{-6}^{+7}$, and $6_{-3}^{+4}$ for LISA, Taiji, and Tianqin, respectively, which is  consistent with the prediction in \cite{2022ApJ...937..118W}, despite different BNS populations from different CBPS models. The uncertainties here show the $10\%$ and $90\%$ quantiles among the $10$ random simulation realizations.

\section{Results}
\label{sec:results}

We present our main results in this Section, first for BNSs in the MW and then for BNSs in LMC and SMC.

\subsection{MW}
\label{subsec:MW}

\begin{table}
\caption{
Estimates for the detection rate of insprialing BNSs in the MW by LISAmax, Folkner, and eASTROD, assuming $T_{\rm obs}=5$\,yr or $10$\,yr. The uncertainties listed here represent the $10\%$ and $90\%$ quantiles among the 10 random simulation realizations.
}
%
\label{tab:t2}
\centering
\begin{tabular}{lcc}  \hline
Detector  & $T_{\rm obs}$\,(yrs) & $N_{\rm MW}$  \\ \hline
\multirow{2}{*}{LISAmax}  & 5 & $564_{-29}^{+25}$ \\
                          & 10 & $833_{-40}^{+35}$ \\
\multirow{2}{*}{Folkner}  & 5 & $816_{-34}^{+32}$ \\
                          & 10 & $1305_{-52}^{+43}$    \\
\multirow{2}{*}{eASTROD}  & 5 & $836_{-42}^{+31}$ \\
                          & 10 & $1331_{-54}^{+43}$ \\ \hline
\end{tabular}
\end{table}

Table~\ref{tab:t2} shows the estimated detection rate of inspiralling BNS in MW with LISAmax, Folkner and eASTROD respectively. The uncertainties here represent the $10\%$ and $90\%$ quantiles among the 10 random simulation realizations. The overall detection rate for MW inspiralling BNS are $N_{\rm MW}\sim 520-1370$, depending on the mission time $T_{\rm obs}$ and GW detectors.  It is not hard to understand that the longer the mission time $T_{\rm obs}$, the larger the SNR. As for quasi-monochromatic GW sources (most of inspiralling BNS considered), the SNR $\rho$ is proportional to $\sqrt{T_{\rm obs}}$ \cite{2015CQGra..32a5014M}.  {Note that while $\rho \propto \sqrt{T_{\rm obs}}$ for individual sources, the total detection number $N$ decreases non-linearly with decreasing $\sqrt{T_{\rm obs}}$ due to the selection effect. For a  short observation period of $T_{\rm obs}=1$\,day, the SNR of a typical BNS would be reduced by a factor of $\sim 60$ compared to that from the mission with whole observation time of $10$-year, resulting in a very low detection rate of $N_{\rm MW}\lesssim 0.5$. As an interesting remark, we find that $N_{\rm MW}$ is about $\sim 6$ and $\sim 33$ with $T_{\rm obs}=1~\rm week$ and $1~\rm month$ respectively. 
}

Besides, it can be seen that the predicted $N_{\rm MW}$ for Folkner and eASTROD are about $\sim 1.5$ times higher than that for LISAmax, which can be explained by their very different power spectral density $S_{\rm n}(f)$. 
{Folkner and eASTROD are substantially more sensitive than LISAmax at frequencies lower than $\sim 4\times 10^{-4}\rm Hz$, while most ($\sim 99\%$) of mock inspirals possess peak GW frequency $f_{\rm p}$ at these lower frequency intervals.} Here the peak GW frequency is defined as the frequency the binaries emitting most power, which can be estimated by \cite{2022ApJ...924..102R},
\begin{equation}
f_{\rm p}=\frac{\sqrt{G(m_1+m_2)}(1+e)^{\gamma}}{\pi [a(1-e^2)]^{3/2}},
\end{equation}
where $\gamma=1.1954$. For example, for a BNS inspiral sample with $e\sim 0.6$ and $P_{\rm orb}\sim 1$\,days, the peak GW frequency lies at $f_{\rm p}\sim  10^{-4}$\,Hz. Thus, Folkner and eASTROD are more sensitive to sources with these parameters and thus may detect more MW inspiralling BNS than LISAmax.

On the contrary, as for frequencies higher than $\sim 10^{-3}$\,Hz, LISAmax is comparably more powerful than the other two detectors, which enhances its capability on detecting highly-eccentric sources. For example, for a system with $e\sim 0.90$ and $P_{\rm orb}\sim 0.5$\,days placing at typical distance of $d_{\rm L}\sim 11.8$\,kpc, the peak GW frequency lies at $f_{\rm p}\sim  1.2\times 10^{-3}$\,Hz (see also Figure~\ref{fig:f3}). Thus, its SNR of LISAmax is $\rho_{\rm L}\sim 40$ for $T_{\rm obs}=10$\,yr, which is about $\sim 2$ times larger that that of Folkner and eASTROD, i.e., $\rho_{\rm F}\sim 16$ and $\rho_{\rm E}\sim 21$.  In total, we find that LISAmax can detect about $\sim 23$ inspiralling BNS with $e>0.90$, while the detection numbers for Folkner and eASTROD are $\sim 17$ and $20$, respectively.

\begin{figure}
\centering
\includegraphics[width=1.0\columnwidth]{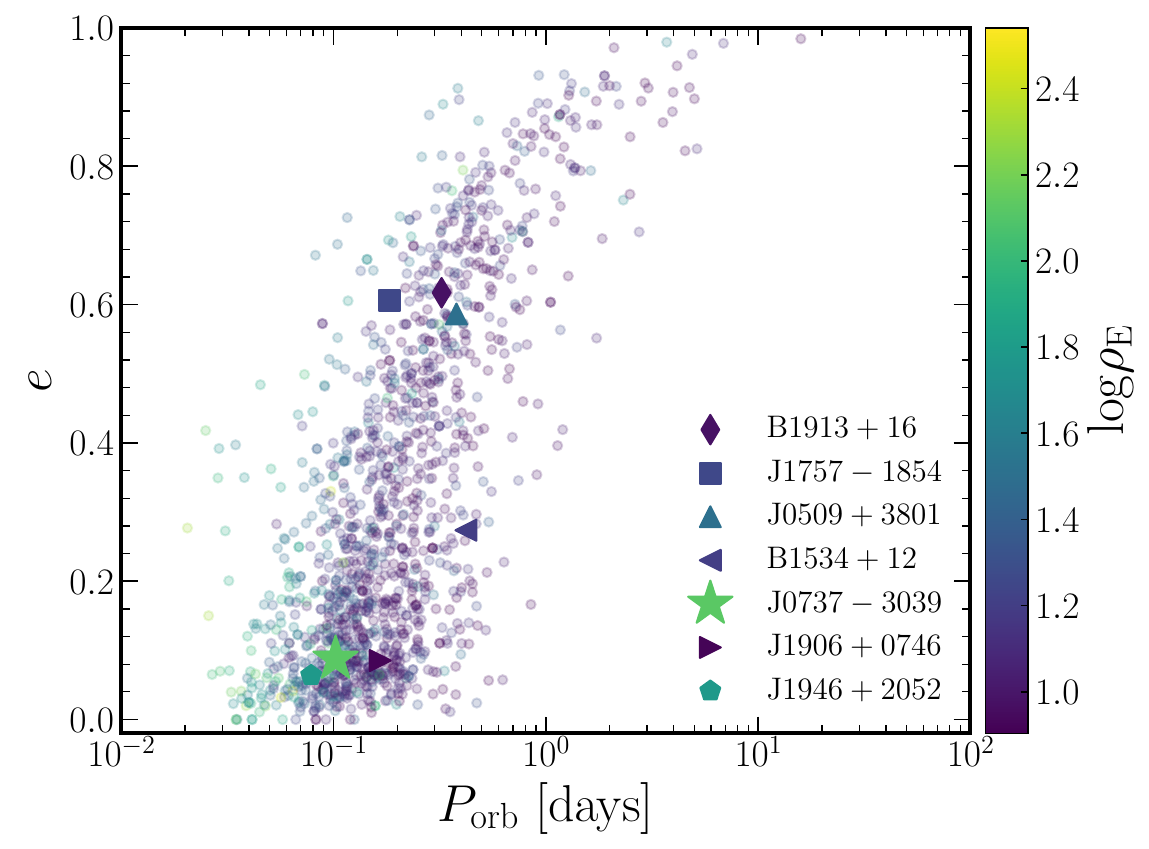}
\caption{The orbital period-eccentricity $P_{\rm orb}-e$ distribution of mock MW inspiralling BNS (filled circles) detected by eASTROD with $T_{\rm obs}=10$\,yrs and their color show the logrithemic SNR $\log \rho_{\rm E}$. Other symbols (as labelled at the bottom right) show seven real MW field BNS systems observed by radio telescopes that can be detected by eASTROD, i.e., $\rm B1913+16$ (diamond), $\rm J1757-1854$ (square), $\rm J0509+3801$ (up triangle), $\rm B1534+12$ (left triangle), $\rm J1906+0746$ (right triangle) and  $\rm J1947+2052$ (pentagon), and $\rm J0737-3039$. The star symbol represents $\rm J0737-3039$ with the largest SNR of $\sim 100$.
}
%
\label{fig:f4}
\end{figure}

\begin{table*}[htbp]
\centering
\caption{Parameters of MW field inspiralling BNS measured by pulsar timing techiniques \footnote{see \url{https://www.atnf.csiro.au/research/pulsar/psrcat/} and \url{http://zmtt.bao.ac.cn/GPPS/}.}. The 2nd-8th columns show the PSR names, luminosity distances determined from dispersion measure (DM) $d_{\rm L}$, orbital periods $P_{\rm orb}$, eccentricities $e$, primary masses $m_1$, secondary masses $m_2$ and total masses $M$ respectively. The 9th-11th columns show the estimated SNR $(\rho_{\rm L}, \rho_{\rm F}, \rho_{\rm E})$ with LISAmax, Folkner and eASTROD assuming mission time $T_{\rm obs}=10$\,yr. Notably, we assume $m_1=m_2$ if only total mass $M$ is provided and $m_1=m_2=1.35M_{\odot}$ if total mass is also undetermined from pulsar observation. The BNSs with * are systems possessing $\rho>8$ at least for one detector.}
\label{tab:t3}
\begin{tabular}{c c c c c c c c c c c}
\hline
No. & PSR  & $d_{\rm L}~[\mathrm{kpc}]$ &$P_{\mathrm{orb}}~[ \mathrm{days}]$  & $e$ & $m_1~[M_\odot]$ & $m_2~[M_\odot]$ & $M~[M_\odot]$ & $\rho_{\rm L}$ & $\rho_{\rm F}$ & $\rho_{\rm E}$ \\
\hline
1 &  B1913+16*   & 9.0 & 0.323 & 0.6171 & 1.44 & 1.389  &2.713& 8.24 & 9.80 & 9.86 \\
2 &  J1757-1854*   & 11.5 & 0.1835 & 0.6058 & 1.338 & 1.395  &2.733& 18.9 & 17.4 & 18.7 \\
3 &  J0509+3801*   & 1.6 & 0.3796 & 0.5864 & 1.34 & 1.46  &2.805& 24.2 & 32.9 & 32.8 \\
4 &  B1534+12*   & 0.9 & 0.4207 & 0.2737 & 1.333 & 1.346  &2.678& 4.58 & 14.7 & 16.0 \\
5 &  J1518+4904   & 1.0  & 8.634& 0.2495 & 1.41 & 1.31  &2.718 & $6.8$$\times$$ 10^{-5}$ & $5.1$$\times$$ 10^{-3}$ & $7.8$$\times $$10^{-3}$  \\
6 &  J1756-2251   & 2.8 & 0.3196 & 0.1806 & 1.341 & 1.23  & 2.570& 2.04 & 6.40 & 6.88 \\
7 &  J1829+2456   & 0.9 & 1.176 & 0.1391 & 1.306 & 1.299  &2.59& 0.05 & 0.83 & 1.03 \\
8 &  J0453+1559   & 0.5 & 4.0725 & 0.1125 & 1.559 & 1.174  & 2.734 & $9.4$$\times$$ 10^{-4}$ & $5.1$$\times $$10^{-2}$& $7.1$$\times$$ 10^{-2}$  \\
9 &  J1913+1102   & 7.1 & 0.2063 & 0.0895 & 1.62 & 1.27  & 2.88 & 2.8 & 6.8 & 6.9 \\
10 &  J0737-3039*  & 1.1 & 0.1023 & 0.0878 & 1.338 & 1.249  & 2.587 &106.3 & 138.9 & 135.9 \\
11 &  J1906+0746*   & 7.4 & 0.166 & 0.0853 & 1.291 & 1.322  & 2.613 & 4.48 & 8.46 & 8.45 \\
12 &  J2150+3427   & 4.8 & 10.5921 & 0.6015 & $<$1.65 & $>$0.94  & 2.59 & $1.6$$\times$$ 10^{-4}$ & $4.9$$\times$$ 10^{-3}$ & $6.5$$\times$$ 10^{-3}$ \\
13 &  J1930-1852   & 2.0 & 45.06 & 0.3989 & $<$1.32 & $>$1.30  & 2.59 & $2.1$$\times $$10^{-9}$ & $5.6$$\times$$ 10^{-6}$ & $5.4$$\times$$ 10^{-4}$ \\
14 &  J1901+0658   & 3.8 & 14.4548 & 0.3662 & - & -  & 2.79 & $6.8$$\times $$10^{-6}$ & $5.9$$\times$$ 10^{-4}$ & $1.1$$\times $$10^{-3}$ \\
15 &  J1759+5036   & 0.5 & 2.043 & 0.3083 & $<$1.92 & $>$0.70  & 2.62 & $3.5$$\times $$10^{-2}$ & 0.58 &0.74 \\
16 &  J1018-1523   & 1.1 & 8.984 & 0.2277 & -  &  $>$1.16  & 2.3 & $3.5$$\times$$ 10^{-5}$ & $2.9$$\times$$ 10^{-3}$ & $4.6$$\times$$ 10^{-3}$ \\
17 &  J1411+2551   & 1.1 & 2.6159 & 0.1699 & $<$1.62 & $>$0.92  & 2.538 & $2.6$$\times$$ 10^{-3}$ & $8.4$$\times$$ 10^{-2}$ & 0.11 \\
18 &  J1325-6253   & 5.4 & 1.8156 & 0.0640 & $<$1.59 & $>$0.98  & 2.57 & $1.2$$\times$$ 10^{-3}$ & $3.7$$\times $$10^{-2}$ & $4.7$$\times $$10^{-2}$ \\
19 &  J1946+2052*   & 3.5 & 0.0785 & 0.0639 & $<$1.31 & $>$1.18  & 2.50 & 56.4 & 66.3 & 65.2 \\
20 & J1753-2240    & 3.0  & 13.6376  & 0.3036  & - &  $>$0.4875  &  - & $6.2$$\times$$ 10^{-6}$ & $6.0$$\times$$ 10^{-4}$ & $1.3$$\times$$ 10^{-3}$  \\
21 & J0528+3529    & 1.9  & 11.7262  & 0.2901  & - &  -  &  - & $1.5$$\times$$ 10^{-5}$ & $1.3$$\times$$ 10^{-3}$ & $2.0$$\times $$10^{-3}$  \\
22 & J1844-0128    &  5.8 & 10.6003  & 0.2349  & - &  -  &  - & $5.0$$\times$$ 10^{-6}$ & $4.6$$\times$$ 10^{-4}$ & $8.5$$\times$$ 10^{-4}$   \\
23 & J1755-2550    &  4.9 & 9.6963  & 0.0894  & - &  $>$0.40  &  - & $3.6$$\times$$ 10^{-6}$ & $4.5$$\times$$ 10^{-4}$ & $9.4$$\times$$ 10^{-4}$   \\
24 & J1208-5936    &  8.5 & 0.631566177  & 0.3480  & 1.26 &  1.32  &  2.568 & 0.19 & 0.76 & 0.86  \\
25 & J1811-1736    &  4.4 & 18.7791691  & 0.8280  & $<$1.64 &  $>$0.93   &  2.57 & $1.3$$\times $$10^{-3}$  & $1.8$$\times$$ 10^{-2}$ & $2.3$$\times $$10^{-2}$  \\
26 & J1155-6529    &  0.7 & 3.67  & 0.2600  & - &  $>$1.27  &  - & $2.3$$\times$$ 10^{-3}$ & $7.5$$\times$$ 10^{-2}$ & $0.10$  \\
27 & J1846-0513    &  5.2   & 0.613021448  & 0.2086 &- &  $>$1.28  & - & 0.15 & 0.93 & 1.08  \\ \hline
\end{tabular}
\end{table*}

Figure~\ref{fig:f4} shows the orbital period-eccentricity $P_{\rm orb}-e$ distribution of mock MW inspiralling BNS detectable by eASTROD and the corresponding colors show their expected SNR $\log \rho_{\rm E}$.  On the one hand, the overall median $P_{\rm orb}$ is about $\sim 0.18$\,days, significantly lower than the intrinsic $\sim 10$\,days. On the other hand, the median value of eccentricity is about $e\sim 0.26$, significantly lower than the intrinsic $\sim 0.44$. These can be explained by the GW detection selection effect, i.e., systems with shorter $P_{\rm orb}$ may have larger SNR but more likely to circularize since formation. We note here that a few BNSs with $e$ close to one and $P_{\rm orb}$ larger than $1$\,day are also detectable, suggesting that the sub-mHz observations may provide important information about the formation of extremely eccentric BNSs (due to large kicks). In addition, we find that a fraction of $\sim 70\%$ detectable events are located within the Galactic latitude $\rm |gb|<10^{\circ}$, corresponding to the thin disk, while the rest $30\%$ are located at larger $\rm |gb|$, away from the thin disk.

\begin{figure}
\centering
\includegraphics[width=1.0\columnwidth]{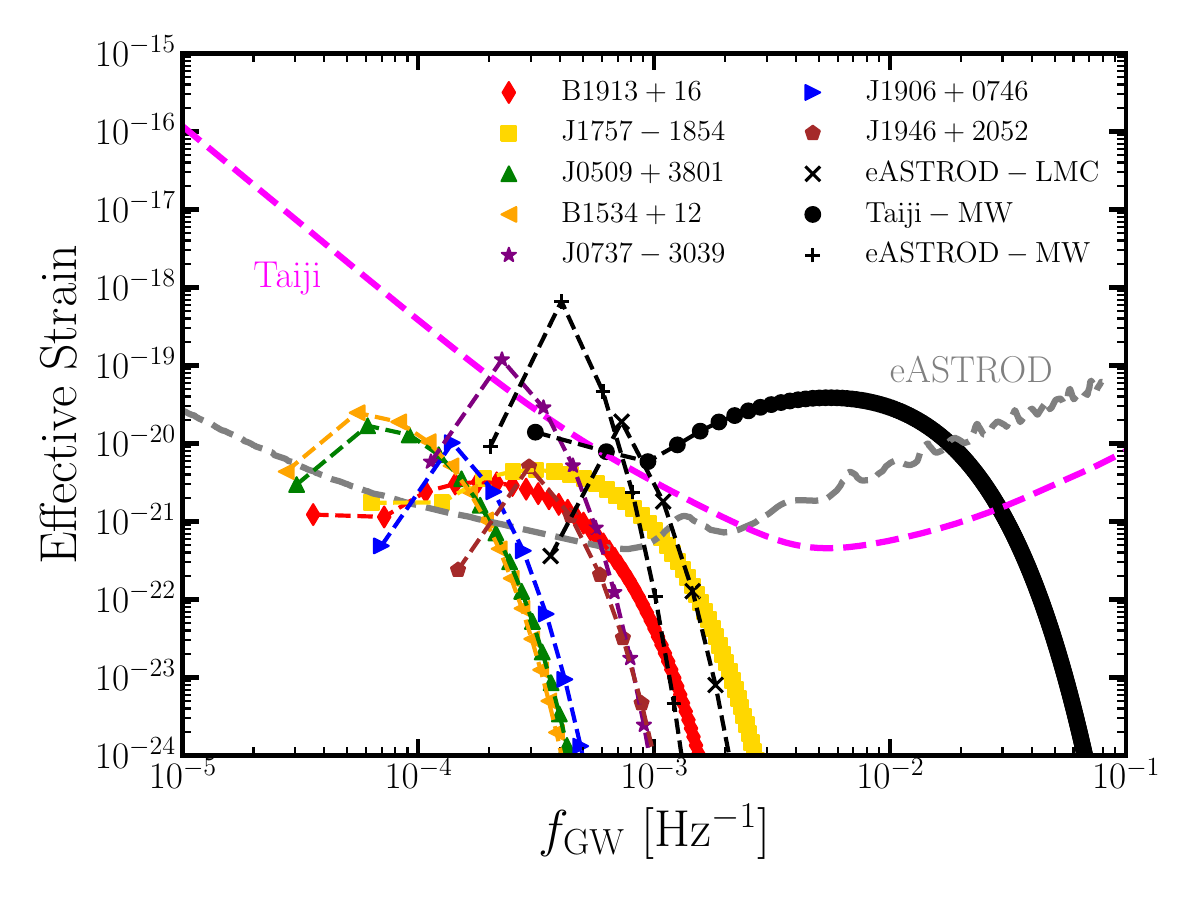}
\caption{The effective strain $h_{{\rm eff},n}$ of seven real MW Inspiralling BNS systems with pulsar components detectable by eASTROD assuming $T_{\rm obs}=10$\,yr. The markers are the same as in Fig.~\ref{fig:f4}. The magenta and grey lines show the characteristic noise strain $h_{\rm noise}$ of mHz GW detector Taiji and sub-mHz detector eASTROD respectively. For comparison, dots, plus, and cross symbols show $h_{{\rm eff},n}$ of the mock inspiralling BNS with the largest SNR among those detectable by Taiji in MW, eASTROD in LMC, and MW, respectively.
}
%
\label{fig:f5}
\end{figure}

For comparison, we also estimate the SNRs for $27$ real MW field BNS systems with pulsar componnent observed by radio telescopes, with well measured parameters from pulsar timing techniques. Their names and properties are listed in Table~\ref{tab:t3}. Among these systems, we find seven, namely, $\rm B1913+16$ , $\rm J1757-1854$ , $\rm J0509+3801$ , $\rm B1534+12$, $\rm J0737-3039$,  $\rm J1906-0736$, and $\rm J1947+2052$, may be detectable by sub-mHz GW detectors. Their properties are also plotted in Figure~\ref{fig:f4} and are well-within the predicted region by our simulations. Figure~\ref{fig:f5} shows their effective strain $h_{{\rm eff},n}$ and one may observe clearly that these systems are rather hard to be detected by mHz GW detectors like Taiji, but may be viewed as nice verification source for sub-mHz detectors. Notably, owing to its small distance $d_{\rm L}\sim 1.1$\,kpc and short orbital period $P_{\rm orb}\sim 0.1$\,days, $\rm J0737-3039$ is expected to have an extremely high SNR, i.e., $(\rho_{\rm L},\rho_{\rm F}, \rho_{\rm E})=(106.3,138.9,135.9)$, making it a very plausible target for detection capability verification of the sub-mHz GW detectors. 

\subsection{LMC and SMC}

\begin{figure}
\centering
\includegraphics[width=1.00\columnwidth]{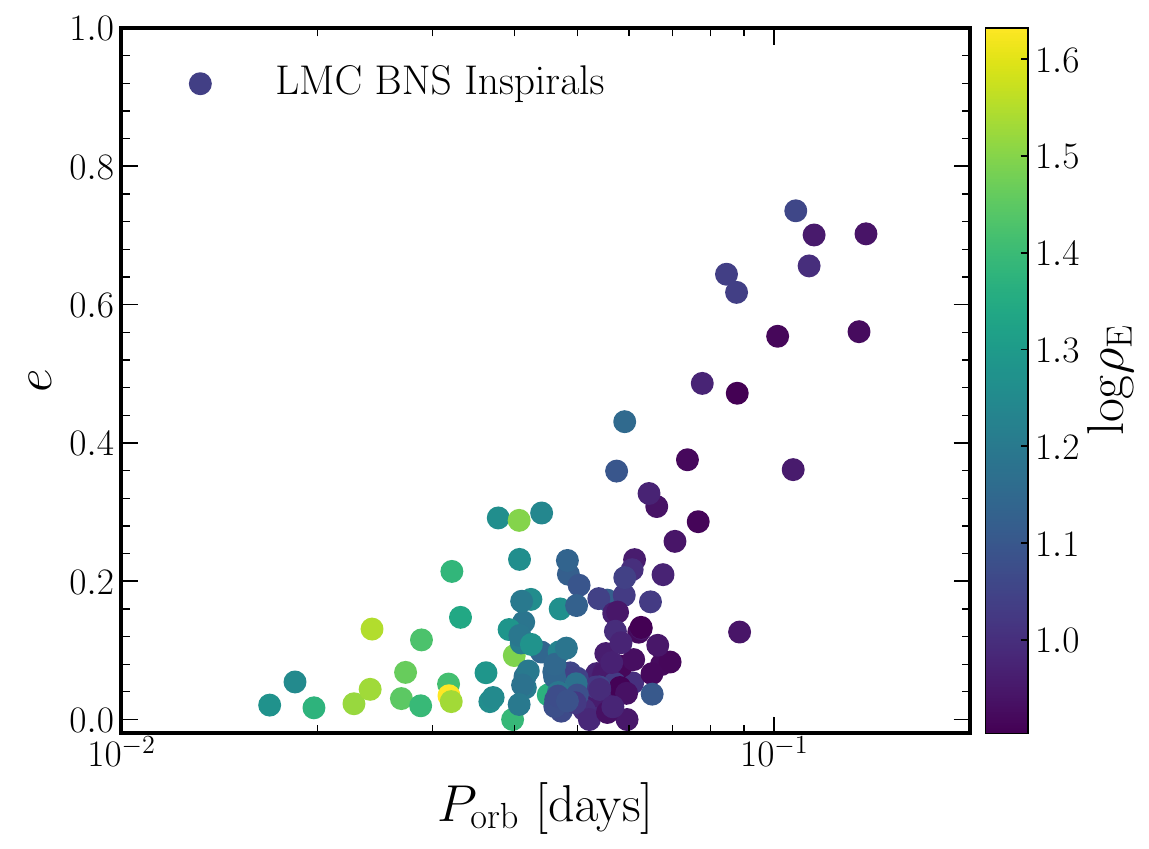}
\caption{The orbital period-eccentricity $P_{\rm orb}-e$ distribution of mock LMC inspiralling BNSs (in 10 realizations) detected by eASTROD with $T_{\rm obs}=10$\,yrs and their color show the logrithemic SNR $\log \rho_{\rm E}$.
}
%
\label{fig:f6}
\end{figure}

For LMC, the predicted detection number of inspiralling BNS of LISAmax, Folkner, and eASTROD are $N_{\rm LMC}\sim 7_{-3}^{+4}$, $6_{-3}^{+3}$, and $7_{-3}^{+2}$ or $\sim 13_{-4}^{+5}$, $11_{-4}^{+5}$, and $13_{-5}^{+5}$, respectively, for $T_{\rm obs}=5$\,yr or $10$\,yr. In Figure~\ref{fig:f5}, for illustration, we plot the effective strain $h_{\rm eff}$ of the system with the largest SNR $\rho_{\rm E}\sim 42.9$ detectable by eASTROD in LMC. Figure~\ref{fig:f6} shows the orbital period-eccentricity $P_{\rm orb}-e$ distribution of mock LMC inspiraling BNSs detected by eASTROD with $T_{\rm obs}=10$\,yrs in $10$ realizations and the corresponding colors show their SNRs. It can be seen that, the largest orbital period $P_{\rm orb}$ among these detectable events is about $\lesssim 0.14$\,days, while the median is about $\lesssim 0.05$\,days, which is substantially smaller than those for the detectable inspiraling BNSs in the MW. This can be explained by the selection effects, i.e., due to the larger distance, the SNRs of the BNSs in LMC are roughly about $\sim d_{\rm L, LMC}/d_{\rm L, MW}\sim 5$ times smaller compared with those in the MW. Thus, only those inspiralling BNS with shorter $P_{\rm orb}$ may be detectable. For the same reason in the MW case, the median eccentricity of detectable inspiraling BNSs is about $e\sim 0.08$ and only a fraction of $\sim 6.3\%$ events possess $e>0.5$.

For SMC, the detection prospect is far more pessimistic with $N_{\rm SMC}\sim 0.6-1$,  which is significantly smaller than that for LMC sources. This is due to their lower star formation rate, the total number of systems with short orbital periods or high eccentricities are $\sim 10$ times smaller than that for LMC. Here we should emphasize that the value of predicted detection number is so small and hence should not be regarded as accurate since it may be largely dominated by the Poisson noise of small number statistics. Nevertheless, we may robustly obtain the message that the detection of inspiraling BNSs by the proposed sub-mHz GW detectors in SMC are significantly challenging.

\section{Conclusions and Discussions}
\label{sec:con}

In this work, we assess the detection capability of future sub-mHz space-borne GW detectors on the inspiralling BNS in MW, LMC and SMC. By simulating mock inspiralling BNS with various parameters through CBPS and galactic dynamics calculation, we further estimate their SNR with LISAmax, Folkner and eASTROD respectively. Our main conclusions are summarized as follows.
\begin{itemize}
    \item The predicted detection number of MW BNS inspiral $N_{\rm MW}$ is about $\sim 520-900$ for LISAmax and $\sim 780-1370$ for Folkner and eASTROD, depending on $T_{\rm obs}$. 
    \item LISAmax may detect more highly eccentric ($e>0.90$) MW BNS systems than Folkner and eASTROD, for its higher sensitivity at higher frequencies $f\gtrsim 10^{-3}$\,Hz.  
    \item  Seven BNS systems discovered by pulsar radio observation may be used as verification of detectability of sub-mHz GW detectors. Among which,  $\rm J0737-3039$ possess SNR of $\rho\sim 100$, making it most plausible.
    \item The predicted detection number in LMC is about $\sim 4-18$ for sub-mHz detectors depending on $T_{\rm obs}$, while detecting inspiralling BNS in SMC is challenging due to their lower abundance and larger distances.
\end{itemize}
 
We note here that there are many complexities one may need to take into account to make a more robust and detailed investigation. First, the mock inspiraling BNS are generated via the CBPS method, which may suffer many uncertainties. In this work, we adopt the model presented in \cite{2025MNRAS.544L..89C}, which is favored by combined Galactic and GW observations; nevertheless, substantial uncertainties remain in the predicted orbital period-eccentricity ($P_{\rm orb}-e$) distribution because of complex binary evolution processes and limited observational data. Analogous to the BNS detection rate assuming different CBPS models estimated by \cite{2022ApJ...937..118W} for LISA, we expect the our estimation on the detection rate of sub-Hz GW detectors may also vary by no more than an order of magnitude similarly. 

{Second, we optimistically follow \cite{2025PhRvD.112h4045W} and \cite{2023PhRvD.107l4022W} in this work to assume that the confusion noise (i.e., foreground) introduced by Galactic DWDs can be accurately modeled and thus may be suppressed significantly. Assuming the most pessimistic circumstance, i.e., the foreground cannot be modeled and subtracted at all, we find that the detection number of MW inspiraling BNSs may reduce to $\lesssim 20$. Therefore, we emphasize that the elimination of the DWD confusion noise is crucial for detecting nearby insprialing BNSs. }

Third, only if the SNRs of at least two loudest harmonics $\rho_a$ and $\rho_b$ are larger than $8$ \cite{2016MNRAS.460L...1S}, one may be able to determine the eccentricity of the detected BNSs from the GW signal with an approximate fractional precision of $\Delta e/e\sim (1/\rho^2_{a}+1/\rho^2_{b})^{1/2}$ \cite{2021MNRAS.502.5576K, 2022ApJ...937..118W}. Among the detectable mock Galactic inspiraling BNSs, we find that only a fraction of $\sim 30\%$ may satisfy the above criteria and their corresponding fractional eccentricity uncertainty is $\Delta e/e\sim 9.6_{-5.9}^{+5.7}\%$, where upper and lower values show the $10\%$ and $90\%$ quantiles among these samples. 

\section*{acknowledgement}
We thank Dr. Zonglin Yang for his help in sorting the MW BNS parameters. This work is partly supported by the National Natural Science Foundation of China (grant nos. 12533009, and 12273050),
the Strategic Priority Program of the Chinese Academy of Sciences (grant no. XDB0550300), the National Key program for Science and Technology Research and Development (grant Nos. 2020YFC2201400 and 2022YFC2205201), the National Astronomical Observatory of China (grant no. E4TG660101), and the Postdoctoral Fellowship Program of CPSF under Grant Number GZB20250735 (ZC).

\bibliographystyle{apsrev4-2}  %
\bibliography{refs}

\end{document}